\documentclass[10pt, conference]{IEEEtran}

\usepackage[T1]{fontenc}
\usepackage{amsmath}
\usepackage{textalpha}
\usepackage[utf8]{inputenc}
\usepackage{interval}
\usepackage[]{hyperref}
\usepackage{booktabs}
\usepackage{tabularx}
\usepackage{multirow}
\usepackage{scalerel}
\usepackage{color}
\usepackage{pgfplots}
\usepackage[ruled,vlined]{algorithm2e}
\usepackage{wasysym}
\pgfplotsset{width=8cm,compat=1.8}
\usepackage{caption}
\usepackage{subcaption}
\usepackage[noabbrev,capitalize]{cleveref}
\usepackage{xcolor}
\usepackage{soul}
\usepackage{siunitx}
\usepackage[htt]{hyphenat}
\usepackage{pgfplots}
\usepackage[noadjust]{cite}

\ifdefined\final
\newcommand{\sout}[1]{}

\newcommand{\todo}[1]{}
\newcommand{\mynote}[1]{}
\newcommand{\ml}[1]{}
\newcommand{\ali}[1]{}

\newcommand{\todoeurosnp}[1]{}
\else
\usepackage[normalem]{ulem}

\newcommand{\todo}[1]{\textcolor{red}{\emph{\textbf{[TODO: }#1\textbf{]}}}}

\newcommand{\mynote}[1]{\textcolor{blue}{\emph{\textbf{NOTE: }#1}}}
\newcommand{\ml}[1]{\textcolor{green}{\emph{\textbf{NOTE (ML): }#1}}}

\newcommand{\todoeurosnp}[1]{\textcolor{orange}{\emph{\textbf{[TODO: }#1\textbf{]}}}}
\newcommand{\ali}[1]{\textcolor{violet}{\emph{\textbf{AT: }#1}}}
\fi

\newcommand{\system}{IREC}

\usepackage{xifthen}
\usepackage[breakable]{tcolorbox}
\tcbsetforeverylayer{autoparskip}
\definecolor{lightgray}{gray}{0.92}
\tcbuselibrary{breakable}

\newcommand{\greybox}[2][]{%
    \begin{tcolorbox}[breakable, colback=lightgray, colframe=lightgray, left=1pt,right=1pt,top=0pt,bottom=0pt]%
        \ifthenelse{\isempty{#1}}{}{\textbf{#1.}\par}%
        #2%
    \end{tcolorbox}%
}

\crefformat{section}{\S#2#1#3} 
\crefformat{subsection}{\S#2#1#3}
\crefformat{subsubsection}{\S#2#1#3}
\crefrangeformat{section}{\S\S#3#1#4 to~#5#2#6}
\crefmultiformat{section}{\S\S#2#1#3}{ and~#2#1#3}{, #2#1#3}{ and~#2#1#3}

\makeatletter
\newcommand{\linebreakand}{%
  \end{@IEEEauthorhalign}
  \hfill\mbox{}\par
  \mbox{}\hfill\begin{@IEEEauthorhalign}
}
\makeatother

\begin{document}

\newcommand{\systemname}[0]{IREC}
\newcommand{\staticinfo}[0]{\texttt{StaticInfoExtension}}

\newcommand{\added}[1]{\textcolor{blue}{\emph{\textbf{ }#1}}}

\renewcommand{\paragraph}[1]{\vspace{1pt}\noindent\textbf{#1.}}


\title{Inter-Domain Routing with Extensible Criteria}



\author{
\IEEEauthorblockN{Seyedali Tabaeiaghdaei}
\IEEEauthorblockA{ETH Zurich}
\and
\IEEEauthorblockN{Marc Wyss}
\IEEEauthorblockA{ETH Zurich}
\and
\IEEEauthorblockN{Giacomo Giuliari}
\IEEEauthorblockA{Mysten Labs\IEEEauthorrefmark{1}}
\and
\IEEEauthorblockN{Jelte van Bommel}
\IEEEauthorblockA{ETH Zurich}
\and 
\IEEEauthorblockN{Ahad N. Zehmakan}
\IEEEauthorblockA{Australian National University\\ Anapaya Systems AG}
\linebreakand 
\IEEEauthorblockN{Adrian Perrig}
\IEEEauthorblockA{ETH Zurich}
}

\maketitle

\begingroup\renewcommand\thefootnote{*}
\footnotetext{The work done when affiliated to ETH Zurich.}

\thispagestyle{plain}
\pagestyle{plain}

\begin{abstract}
With the rapid evolution and diversification of Internet applications, their communication-quality criteria are continuously evolving. To globally optimize communication quality, the Internet's control plane thus needs to optimize inter-domain paths on diverse criteria, and should provide flexibility for adding new criteria or modifying  existing ones. However, existing inter-domain routing protocols and proposals satisfy these requirements at best to a limited degree.

We propose \system{}, an inter-domain routing architecture that enables multi-criteria path optimization with extensible criteria through parallel execution and real-time addition of independent routing algorithms, together with the possibility for end domains to express their desired criteria to the control plane. We show \system{}'s viability by implementing it on a global testbed, and use simulations on a realistic Internet topology to demonstrate \system{}'s potential for path optimization in real-world deployments.


\end{abstract}
\section{Introduction}\label{sec:intro}

\begin{sloppypar}
Operators of large wide-area networks~(WANs) have long realized the importance of \emph{intra-domain} path optimization based on demand and traffic characteristics. 
In these networks, often based on software-defined networking (SDN), centralized controllers (i) monitor the state of the network and analyze the specific requirements of traffic flows; (ii) optimize the network paths and assign them to flows according to such requirements; and (iii) constantly and rapidly update to accommodate new applications with different path-optimization criteria~\cite{Jain:2013:B4, Yap:2017:espresso}. These networks thus provide a flexible and performant platform, where traditional applications thrive and new ones can evolve, such as cloud-based gaming~\cite{forbes-online-gaming} or holographic communication~\cite{Ha:2022:holographic}.
%
In contrast, the scale and complexity of \emph{inter-domain} routing has prevented analogous advancements on the Internet to this point.
Today, Internet path optimization is limited to coarse-grained strategies, such as selective prefix announcements~\cite{Cardona:2016:bgp_filtering} and BGP communities~\cite{Borkenhagen:2019:bgp_communities}; only the largest Internet service providers (ISPs) and content-distribution networks (CDNs) can afford to deploy anycast routing to boost last-mile performance.

Path-aware networks (PANs) emerged as a new paradigm to introduce more granular inter-domain path optimization.
PAN architectures~\cite{Segment_Routing, Chuat2022SCION,Pathlets} expose multiple paths to applications at the endpoints, allowing them to select the best path based on application requirements. 
To further inform the endpoints' decision and increase the potential for optimization, their routing messages may contain multiple performance metrics, such as latency and bandwidth.
%
However, while performance-aware path selection at the endpoints can significantly improve communication quality compared to today's single-path Internet~\cite{conext2021deployment}, it still cannot compete with the optimization granularity of intra-domain networks. 
Unlike controllers in a centralized WAN, 
autonomous systems (ASes) in the Internet cannot fully optimize their routing decisions to satisfy their customer's needs: Fundamentally, ASes lack full visibility into the inter-domain topology and are oblivious to the requirements of endpoints outside their domains. 
Furthermore, the distributed nature of inter-domain routing significantly complicates and delays upgrades to the routing infrastructure~\cite{Wirtgen:2023:xbgp}---compared to the almost instantaneous upgradability of SDN networks---thus limiting extensibility and hindering innovation.

Our goal is, therefore, to design a control-plane architecture that enables fine-grained Internet path optimization according to diverse criteria, and can continuously evolve to accommodate the ever-changing needs of new applications.
Despite the complexity of this goal, we can tackle it thanks to the following key intuition. In PANs, the use of paths for forwarding is completely separate from the path-discovery mechanism---unlike, e.g., in BGP, where a routing update can completely change the course of packets in flight. Thus, multiple such path-discovery mechanisms can coexist in parallel without interfering: The result is a wider choice of paths, optimized for different objectives, which endpoints can select from according to their needs. 
%
Our inter-domain routing with extensible criteria (\system{}) architecture builds on this intuition, providing an framework for ASes to deploy independent path-optimization algorithms in parallel (\Cref{sec:principles:parallel}). Each AS can choose which algorithms to deploy, and new algorithms can be dynamically loaded into the system without compromising other live routing processes. 

The bulk of our work is therefore devoted to addressing several scalability challenges that deploying parallel routing algorithms poses for ASes, as well as solving the tension between \emph{global optimization}---that requires a common understanding of optimization metrics and algorithms---and \emph{high extensibility}---which is bound to rapidly change such understanding.
%
Interestingly, these challenges can be overcome by
moving beyond the usual destination-initiated route-update flooding mechanism. 
In \system{}, we introduce two forms of \emph{interactive} routing mechanisms, where ASes do not passively forward routing updates, but request other ASes in the network to optimize path selection towards their goals.
%
The first is \emph{on-demand routing} (\Cref{sec:principles:on_demand}), whereby ASes express their optimization criteria for paths by communicating routing algorithms via routing messages. 
The second is \emph{pull-based routing} (\Cref{sec:principles:pull_based}), which allows traffic sources to request paths with specific characteristics towards a destination---the opposite of traditional destination-generated routing messages.
With these two mechanisms in place, ASes can deploy new routing algorithms in real time without touching the existing infrastructure.

\medskip
\noindent
In summary, we contribute towards a performant and extensible Internet by: 
\begin{itemize}
    \item Designing \system{}, a control-plane architecture for PANs that enables extensible and multi-criteria path optimization;
    \item Implementing and benchmarking \system{} in the open-source codebase~\cite{scionproto} of a PAN, SCION~\cite{Chuat2022SCION} (\Cref{sec:implementation}); and
    \item Evaluating \system{} on a realistic Internet topology using an ns-3~\cite{nsnam_ns-3_nodate} based simulator~\cite{scion-sim}, showing its optimization potentials on multiple criteria (\Cref{sec:simulations}).
\end{itemize}

\end{sloppypar}

\section{Problem Statement}\label{sec:statement}

\definecolor{plotred}{HTML}{fc4f30}
\definecolor{plotblue}{HTML}{008fd5}
\usetikzlibrary{%
  automata, positioning, arrows, arrows.meta, shapes.symbols, shapes.multipart, shapes.geometric,
  decorations.pathreplacing, matrix, backgrounds, calc, chains
}
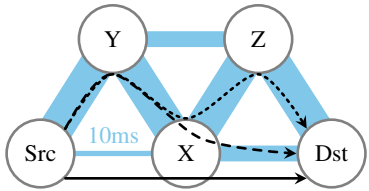
\begin{figure}[t]
\begin{center}
    \small
\begin{tikzpicture}[
    scale=1, transform shape,
    asnode/.style={draw=black!50, fill=white, circle, minimum size=0.9cm, line width=1pt, },
    highbw/.style={draw, line width=13, plotblue!50},
    midbw/.style={draw, line width=6pt, plotblue!50},
    lowbw/.style={draw, line width=2pt, plotblue!50},
    ]
    \node (SRC) [asnode] {Src};
    \node (X) [asnode, right=1cm of SRC] {X};
    \node (DST) [asnode, right=1cm of X] {Dst};
    \node (Y) [asnode, above right=0.86 cm and 0.3 cm of SRC] {Y};
    \node (Z) [asnode, above right=0.86 cm and 0.3 cm of X] {Z};

    \begin{scope}[on background layer]
        \draw [lowbw] (SRC.center) edge node[anchor=south] {10ms} (X.center);
        \draw [highbw] (SRC.center) edge (Y.center);
        \draw [midbw] (X.center) edge (DST.center);
        \draw [midbw] (Y.center) edge (Z.center);
        \draw [highbw] (X.center) edge (Z.center);
        \draw [highbw] (Y.center) edge (X.center);
        \draw [highbw] (Z.center) edge (DST.center);
    \end{scope}

    \draw [black, line cap=round, -stealth, line width=1pt, smooth] plot coordinates {(SRC.south east) (DST.south west)};
        
    \draw [black, line cap=round, -stealth, dashed, line width=1pt, smooth] plot coordinates {(SRC.north east) (Y.south) ($(X.center) + (0.2,0.2)$) (DST.west)}; 
    \draw [black, line cap=round, -stealth, dotted, line width=1pt, smooth] plot coordinates {(SRC.north east) (Y.south) (X.north) (Z.south) (DST.north west)};
\end{tikzpicture}
\end{center}
\caption{Example of extensible multi-criteria path optimization: Each link between ASes (circles) adds 10ms of latency to each path. Links also have different bandwidth, represented by the line thickness. Three paths are highlighted, each representing a different optimization trade-off: The shortest (continuous arrow), the highest-bandwidth (dotted arrow), and the highest-bandwidth path with latency $\leq\SI{30}{\milli\second}$ (dashed arrow).}%
\label{fig:example_criteria}
\end{figure}

\subsection{Motivation}\label{sec:statement:motivation}
\paragraph{Multi-Criteria Path Optimization}\label{sec:statement:motivation:multi_criteria} Network endpoints typically have application-specific communication-quality criteria, for which the control plane should ideally provide optimal paths. However, in the scope of inter-domain routing, despite the diversity of potentially discoverable paths~\cite{conext2021deployment}, paths are usually optimized on a few static criteria such as AS-hop length while respecting business relationships.

\greybox{\emph{Example \#1}: We provide an example of the usefulness of multi-criteria path optimization referencing \cref{fig:example_criteria}. 
Consider two applications running at the source AS: A VoIP client, for which lower communication latency directly results in improved quality of experience, and a file-transfer application, that requires a high-bandwidth path. 
BGP, a single-path routing protocol, would only route the shortest path by number of hops---which in this case also corresponds to the lowest-latency path (continuous arrow). This path is excellent for the VoIP application, but suboptimal for the file-transfer application. The best path for file transfer, i.e, the highest-bandwidth path (dotted arrow), needs to be independently discovered by the control plane. For this to happen, the control plane needs to optimize routing updates for two different metrics.}

\paragraph{Criteria Extensibility}\label{sec:statement:motivation:extensibility} As applications continue to evolve, so do their communication requirements. Additionally, ASes need to continuously adjust their routing decisions as the result of network changes, or routing policy updates. 
Optimizing paths on extensible criteria is thus the ability to modify path optimization criteria according to application or network requirements without interrupting connectivity or disrupting control-plane functionality and with low effort, i.e., requiring no standardization or vendor implementation, and minimal deployment effort by network operators.
However, in current inter-domain routing, such adjustments require standardization efforts, which often unfold over the period of a decade~\cite{Wirtgen:2023:xbgp}, and then require implementation by vendors, and finally deployment by network operators. 
How to shorten decades-long innovation cycles in inter-domain routing and enable support for emerging application requirements is an intriguing research challenge.
The significance of the problem is well underscored by the multitude of research studies, such as XIA~\cite{Naylor:2014:XIA}, Trotsky~\cite{McCauley2019Trotsky}, and xBGP~\cite{Wirtgen:2023:xbgp}, which take significant steps in making  inter-domain routing extensible.

\greybox{\emph{Example \#2}: Consider again the example in \cref{fig:example_criteria}. A new application for live-video streaming is deployed at the source. For this application, the optimal network path provides the highest bandwidth possible \emph{with bounded latency} (e.g., $\leq \SI{30}{\milli\second}$): users want to experience live events without buffering, and without falling behind the action. Therefore, the two paths discovered before are insufficient: The shortest path's bandwidth (continuous arrow) is too low, while the highest-bandwidth path is too long (40ms). Therefore, the control plane in the first example cannot accommodate this requirement and needs to be extended to include the notion of latency bounds. 
Only then the source can discover the only path that satisfies the live-video requirements (dashed arrow).}

\subsection{Desired Properties}\label{sec:statement:properties}
To realize multi-criteria path optimization and extensibility, we seek the following properties in a control-plane design: 
\begin{itemize}
    \item[\textbf{P1}] Ability to find optimal paths for multiple criteria.
    \label{p1}
    \item[\textbf{P2}] Independent selection of criteria by each AS.\label{p2}
    \item[\textbf{P3}] Real-time, low-effort, and interruption-free addition and removal of criteria.\label{p3}
    \item[\textbf{P4}] Ability to optimize paths on the criteria of both source and destination ASes of data packets.\label{p4}
    \item[\textbf{P5}] Providing means of optimality guarantee for any criterion.\label{p5}

\end{itemize}

\subsection{Challenges}\label{sec:statement:challenges}

\paragraph{Scalability}\label{sec:statement:challenges:scalability}
Different network paths may be optimal for different criteria. Thus, finding optimal paths on all criteria requires multipath routing and forwarding. In stateful multipath proposals, the routing and forwarding state at a router grows linearly with the number of paths. Therefore, the limited forwarding and routing memory in routers impedes scalability regarding number of paths, and in turn, the scalability regarding number of optimization criteria. Furthermore, required computational resources for optimizing paths on multiple criteria can limit scalability to large number of criteria. 

\paragraph{Forwarding Stability} Multi-criteria path optimization can  destabilize stateful forwarding by: (i) increasing routing convergence time for discovering multiple paths, and (ii) increasing the probability of path changes on routers because of the increased probability of change in any of multiple performance metrics compared to a single performance metric. Both can lead to transient forwarding loops and blackholes, and thus, disconnectivity. Extensibility can also introduce instability in stateful forwarding: modifying routing decisions can change preferred paths, which can cause transient loops and blackholes.

\paragraph{Optimality vs. Extensibility}
To guarantee connectivity and optimality on any criterion, all ASes need to agree on (i) optimizing that criterion, and (ii) the format of routing messages. Such requirements impede extensibility, as it needs significant effort to ensure all ASes optimize on the same criteria. Therefore, the main challenge is to reconcile extensibility with optimality and connectivity guarantee.


\paragraph{Usability}
The ultimate goal of multi-criteria path optimization is to satisfy traffic flows' requirements. This requires the \emph{data plane} to be aware of these requirements and forward each traffic flow on the most suitable path(s). 
However, such a mechanism is currently missing in the inter-domain data plane. 
For this reason, we build \system{} within the scope of a path-aware network architecture.

\section{The Promise of a PAN Architecture}\label{sec:pan}

The forwarding stability and scalability challenges of the current Internet we highlighted in the previous section stem from two of BGP's inherent characteristic: Stateful inter-domain forwarding, and the tight coupling of control-plane operations and forwarding state. Although many studies have endeavored to tackle them by proposing modifications to BGP~\cite{van_den_schrieck_bgp_2010,walton_dwaltonciscocom_advertisement_2002,bgp_xm_2013,dimr_2013,amir_2012,yamr_2010,mifo_2015,stamp_2008,dbgp_and_bbgp_2009,path_splicing_2008}, they were not able to completely overcome them.
Therefore, we seek an alternative that resolves the BGP's  inherent challenges, and thus provides a reliable foundation for extensible multi-criteria path optimization. 
SCION~\cite{Chuat2022SCION} is an alternative Internet PAN architecture that satisfies all these requirements. In the following, we provide an overview of SCION with a focus on the features that most influence our design.

\paragraph{Data Plane} In SCION, the inter-domain data plane is stateless by virtue of using packet-carried forwarding state, i.e., each packet contains the whole inter-domain forwarding path in its header. The path specifies ingress and egress interfaces to/from every on-path AS. Border routers in each AS forward packets to the next border router according to this information, and therefore do not need to keep inter-domain forwarding state. 

\paragraph{Control Plane} The SCION data and control planes are separated. Dedicated SCION \emph{control services} manage the control plane in each AS. The control services in different ASes collaborate to run the periodic \emph{beaconing} process, which constructs paths by originating and propagating routing messages called path construction beacons (\emph{PCB}s). Similarly to BGP, each AS selects a set of preferred paths to each origin, extends them by adding its own AS hop, and propagates them to neighboring ASes. However, there are fundamental differences to BGP: (i) beaconing is run by servers, instead of routers, (ii) paths are stored in scalable databases, instead of a longest-prefix matching table, (iii) paths are specified per \emph{origin AS}, instead of per prefix, (iv) multiple paths can be advertised per origin AS, (v) paths specify each AS hop in the granularity of ingress and egress interfaces, (vi) each AS signs its hop-information in PCBs, and (vii) all PCBs have validity times set by their origin ASes but limited by a global hard upper bound.  

Each AS hop in a PCB may contain multiple \textit{static info extension} providing information about path performance metrics---e.g, link bandwidth---which can be used for optimizing paths based on different criteria.

\paragraph{Endpoint Path Selection} The control service in each AS registers a subset of the beaconed paths to the path service of its AS. Endpoints in each AS contact their AS' path service and request paths to any destination AS. 
The path service's response contain multiple paths, along with information about their performance metrics. Endpoints can select the paths that best suit their needs and use them for forwarding.

\paragraph{Deployment} SCION is commercially deployed in an BGP-free global network by multiple ISPs ~\cite{conext2021deployment}. This successful incremental deployment has shown the practicality of a BGP alternative in the real world that can co-exist with the BGP-based Internet, without problems of BGP affecting the SCION network. Independent from this commercial deployment, SCION is also deployed in the SCIONLab~\cite{Kwon:2020:SCIONLab} global research network as an overlay, providing researchers with the opportunity to run SCION experiments.

\paragraph{Addressing Challenges} With stateless forwarding and separation of control and data planes, SCION eliminates the forwarding stability and scalability challenges. Thanks to path-awareness, discovered paths can be readily used by end-hosts. Also, with a software-based control plane running on servers, it provides the required foundation of an extensible control plane; like in SDN, ASes do not need to wait for vendor support nor sandardization for to modify in their routing logic in ways that may not be supported by existing hardware or standards.

\paragraph{A New Control-Plane Design} While the current SCION control plane does not satisfy the properties \hyperref[p1]{P1}-\hyperref[p5]{P5} and does not solve all the mentioned challenges in~\Cref{sec:statement:challenges}, it provides the reliable foundation we need to design such a control plane, i.e,~\system{}.
By relying on a deployed network architecture,~\system{} benefits from a promising deployment opportunity in production networks.

\begin{figure*}[t!]
    \label{fig:principles:main_concepts}
    \includegraphics[width=\linewidth]{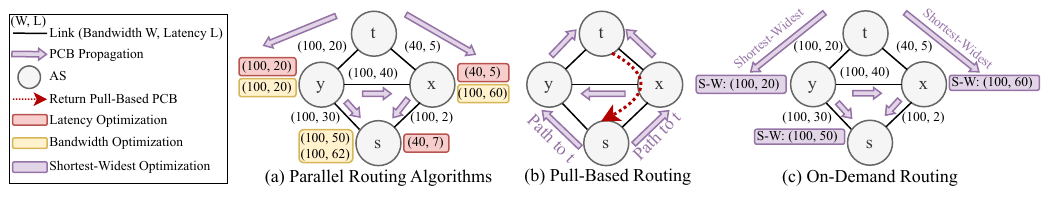}
    \caption{Examples of (a) parallel routing algorithms, where one algorithm optimizes for low latency, and one for high bandwidth,
(b) pull-based routing, in which the origin AS~$s$ initiates beaconing towards a specific target AS~$t$, which ultimately returns a subset of constructed PCBs back to AS~$s$, and (c) on-demand routing, where origin AS~$t$ specifies shortest-widest algorithm, which selects the lowest-latency path among the highest-bandwidth ones, as its criterion in PCBs, using which other ASes optimize paths. \vspace*{-11pt}}
\end{figure*}

\section{\system{}'s Routing Mechanisms} \label{sec:principles}

This section describes \system{}'s mechanisms to achieve multi-criteria path optimization and criteria extensibility.

\subsection{Parallel Routing Algorithms}\label{sec:principles:parallel}
Motivated by the diverse criteria of applications, we propose the concepts of \emph{separate application-wise criteria sets} and \emph{independent optimization} for each such criteria set to achieve multi-criteria path optimization (\hyperref[p1]{P1}).
A criteria set is a subset of all possible criteria across the Internet required by at least one type of application in at least one end domain. 
Independent criteria sets allow low-effort addition of new ones, without requiring any modification to existing ones. 
To allow separate optimization of criteria sets, \system{} enables execution of multiple routing algorithms in parallel and independent of each other, i.e., every algorithm optimizes on one specific criteria set.
Every AS can decide to deploy an instance of an algorithm independently and without any coordination with other participating ASes, facilitating the deployment of new algorithms (\hyperref[p2]{P2},\hyperref[p3]{P3}). Figure~\hyperref[fig:principles:main_concepts]{2a} shows an example of parallel routing algorithms.
Algorithm instances running in different ASes communicate with each other through routing messages (PCBs), where participating ASes extend those messages with the relevant performance metrics, e.g., latency information.
One PCB can thus contain path metadata about multiple performance metrics used by multiple algorithms.

\subsection{Pull-Based Routing}\label{sec:principles:pull_based}

We propose pull-based routing, using which an AS can retrieve additional paths to a \textit{specific target AS} that are not provided by upstream ASes through conventional routing.
Thereby, an AS originates PCBs containing the identifier of the target AS. Any non-target AS receiving these PCBs propagates a subset of them to its neighboring ASes.
This process continues until the PCBs arrive at the target AS, which sends them back to their origin AS.
Pull-based routing is different from conventional routing in two ways: (i) the propagation of routing messages is in the opposite direction of conventional routing protocols; in pull-based routing, an AS initiates the routing process to reach another AS, while in conventional routing, an AS initiates the routing process to be reached by all other ASes, (ii) routing messages specify both end ASes of routing while in conventional routing, only one end is specified. Figure~\hyperref[fig:principles:main_concepts]{2c} further illustrates the idea of pull-based routing.

\subsection{On-Demand Routing}\label{sec:principles:on_demand}
Based on the idea of independent optimization on criteria sets, we introduce on-demand routing to (i) allow end ASes to express their criteria sets to optimize paths accordingly, (ii) allow real-time, automated and interruption-free addition of new criteria sets (\hyperref[p3]{P3}), (iii) ensure the scalability of the system to a large number of criteria sets (\Cref{sec:statement:challenges:scalability}), and (iv) ensure all ASes are running the same algorithm, which guarantees routing consistency and provides the necessary means of discovering optimal paths (\hyperref[p5]{P5}).

An AS can use on-demand routing by originating PCBs and encoding its desired routing algorithm (optimizing on a criteria set) that is to be executed by other ASes in the PCBs, where each PCB can contain at most one algorithm.
The AS can also specify fine-grained constraints in its algorithm, e.g., latency bound or (un)desired geographical areas, ASes, or links.
When the origin AS is the source of data traffic, to optimize paths according to its criteria to a target AS, it needs to use both pull-based and on-demand routing mechanisms together~(\hyperref[p4]{P4}).
When a non-origin AS on the propagation path receives such a PCB, it automatically loads and executes the corresponding algorithm.
The optimality of the discovered paths is guaranteed, since all ASes execute the same algorithm.
%
Importantly, an on-demand algorithm~$A$ from origin AS~$t$ only optimizes on the set of PCBs containing~$A$ and originating from~$t$. 
The main benefit of this approach is the tremendous reduction of the number of PCBs an algorithm has to process, which lowers its execution time and required resources, allowing \system{} to scale to a large number of criteria.
Figure~\hyperref[fig:principles:main_concepts]{2b} provides an example of on-demand routing.
%

\subsection{Flexible Optimization Granularity}\label{sec:principles:granularity}

\begin{figure}[t!]
    \centering
    \includegraphics[width=\linewidth]{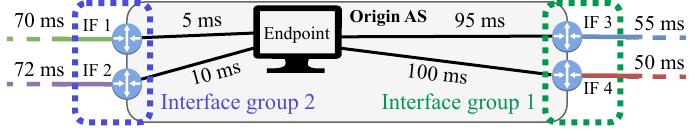}
    \caption{Motivation for interface groups. AS-based latency optimization prefers paths to interface \num{4}, leading to suboptimal latency to endpoints far from that interface.
Interface-based optimization finds one path per interface (four in this case), introducing significant communication and computation overhead in the control plane. By optimizing paths per group of e.g. geographically close interfaces a close-to-optimal low-latency path to the endpoint can be discovered efficiently.}
\label{fig:principles:interface_groups}
\end{figure}

Ultimately, we not only aim at optimizing paths between ASes, but in particular between applications' endpoints. 
However, SCION PCBs specify path origins at the granularity of ASes and their ingress interfaces.
This only allows path optimization per origin AS, i.e., finding optimal paths to every origin AS, or per interface, i.e., finding optimal paths to every interface of every origin AS.
This is a rigid granularity scheme with two extreme options at the two ends of a wide range of possible optimization granularity: AS-based optimization is too coarse-grained to provide optimal end-to-end paths, violating property \hyperref[p5]{P5}, while interface-based optimization is too fine-grained and introduces excessive communication and computation overhead.

To address these problems, we introduce the concept of \emph{interface groups}, representing sets of interfaces, and propose per-interface-group path optimization.
To that end, origin ASes create interface groups based on their preferences and assign each of their interfaces to some of these groups.
For each group, they originate PCBs from all its member interfaces and encode their group ID in the PCBs.
Non-origin ASes then perform path optimization on each interface group separately.
Interface groups allow to flexibly specify the granularity of path optimization; origin ASes can adjust them by changing the interface-to-group-assignments and subsequently originating new PCBs.
\Cref{fig:principles:interface_groups} illustrates the benefits of interface groups.

\subsection{Optimization on Extended Paths}\label{sec:principles:per_egress}

The current SCION routing abstracts away the internal network of each AS, as it only optimizes the received paths.
This abstraction neglects the fact that the AS' internal network can affect performance metrics.
This can cause suboptimality of paths, thus violating property \hyperref[p5]{P5}, because the relative preferences of any two received paths can change when they are propagated to the same interface.\footnote{This translates into non-isotonicity of a criterion on path extension~\cite{Sobrinho:2020:multiple_optimality}.}
%
%
We address this problem by optimizing \emph{extended paths} instead of received paths:
For every egress interface, an AS extends the performance metrics of received inter-domain paths with the performance metrics of intra-domain paths connecting to that interface.
Then, it optimizes paths based on those extended performance metrics.
For example, in ~\Cref{fig:principles:extend_then_optimize}, computing the latency of extended paths at the egress interface before propagating them enables the discovery of optimal paths.

\begin{figure}
    \centering
    \includegraphics[width=1.0\linewidth]{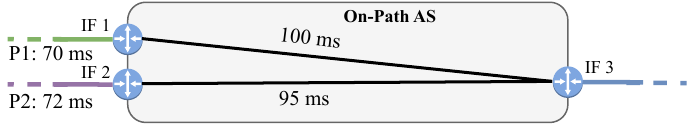}
    \caption{
    Motivation for path extension before optimization (\cref{sec:principles:per_egress}).
    If the on-path AS optimizes received paths P1 and P2 for latency, the AS propagates P1 to interface~3, which is suboptimal because the intra-domain path connecting interfaces 1 and 3 has higher latency than the one connecting interfaces 2 and 3.
    When optimizing on the extended paths, however, the AS propagates P2, achieving optimal latency.
    }
    \label{fig:principles:extend_then_optimize}
\end{figure}

\subsection{PCB Extensions} 
\label{sec:principles:pcb_extensions}
We summarize the new PCB extensions introduced by \system{}'s mechanisms:

\begin{itemize}
    
    \item \textit{Target}, specifying a target AS; at most one per PCB added by the origin AS to use pull-based routing (\Cref{sec:principles:pull_based}),
    \item \textit{Algorithm}, specifying a routing algorithm's identifier and hash of implementation code; at most one per PCB added by the origin AS to use on-demand routing (\Cref{sec:principles:on_demand}), and
    \item \textit{Interface group}, specifying an interface group; at most one per PCB added by the origin AS to use flexible optimization granularity (\Cref{sec:principles:granularity}).

\end{itemize}

\section{\system{}'s Intra-AS Architecture}\label{sec:internal_architecture}
In this section, we describe the different intra-AS components and their interactions required by an AS to implement \system{}'s routing mechanisms described in \cref{sec:principles}.
\system{} is backwards compatible and can be incrementally deployed.


\subsection{Overview}\label{sec:internal_architecture:overview}
There are three main components in \system{}: (i) the \textit{ingress gateway}, which receives and stores incoming routing messages (PCBs), (ii) \textit{routing algorithm containers} (RACs), which are programs executing routing algorithms to select the best PCBs according to some set of criteria, and (iii) the \textit{egress gateway}, which collects the selected PCBs from the RACs, filters duplicate ones, and propagates them further to neighboring ASes.
The egress gateway is also responsible for originating new PCBs.
\system{} enables parallel routing algorithms through the deployment of the corresponding RACs.
\Cref{fig:system_overview} provides an overview of \system{}'s components.
The components only affect the control- but not the data plane, meaning that neither changes to the SCION packet header nor modifications to the routers' forwarding behavior are necessary.

\begin{figure}
    \centering
    \includegraphics[width=\linewidth]{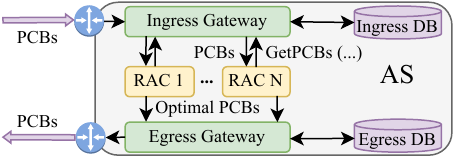}
    \caption{Overview of \system{}'s intra-AS architecture.}
    \label{fig:system_overview}
\end{figure}

\subsection{Ingress Gateway}\label{sec:internal_architecture:ingress}
When receiving a PCB from a neighboring AS, the ingress gateway verifies the included signatures and whether the path constructed by the PCB complies with the local AS' policies.
The ingress gateway then stores the PCB in its \textit{ingress database}.
The gateway periodically removes (soon-to-be) expired PCBs from the database.

\subsection{Routing Algorithm Containers}\label{sec:internal_architecture:rac}
RACs are \system{}'s way of supporting multiple routing algorithms in parallel (\cref{sec:principles:parallel}).
A RAC provides the required environment to run a routing algorithm, and communicates with it using a standardised interface~(\cref{sec:deployment_standardization}).
RACs have access to AS topology information such as latencies between AS interface-pairs, which they can use for the enhanced path optimization described in \cref{sec:principles:per_egress}.
In a typically periodic pattern, a RAC requests PCBs from the ingress gateway, provides them as inputs together with intra-AS topology information to its algorithm, executes the algorithm, and ultimately gets back the set of optimal PCBs.
The PCBs provided as input are specific for an origin AS, as well as interface group and target AS (if available in the PCB and enabled in the RAC configuration); those parameters are not explicitly communicated to the algorithm.
For the optimal set of PCBs returned by the algorithm, the maximally allowed size is configurable for each RAC and egress interface.
%
Subsequently, a RAC sends the selected PCBs, together with the IDs of the egress interfaces for which they are optimized, to the egress gateway.
We distinguish between two RAC types: \textit{static RACs} and \textit{on-demand RACs}.
%

\paragraph{Static}
A static RAC always runs the same algorithm configured by its AS.
Examples are RACs with algorithms optimizing for latency, bandwidth, or number of on-path ASes.

\paragraph{On-Demand}
On-demand RACs are dedicated to run on-demand routing algorithms communicated via PCBs.
The main difference to static RACs is that on-demand RACs thus change their routing algorithm over time.
Because on-demand algorithms are communicated via their IDs (\cref{sec:principles:pcb_extensions}), an on-demand RAC has to fetch the actual executable from the origin AS of the corresponding PCBs; by caching the executable, the RAC only needs to do this once for all PCBs with the same origin AS and algorithm ID.
Note that there are no cyclic dependencies introduced here: the origin AS can always be reached via the path contained in the PCB itself, or alternatively over any other previously discovered path.
The RAC only allows executables up to a certain size limit, and verifies that the hash of the fetched algorithm is equal to the hash included in the PCB.
The hash is computed using a collision-resistant hash function, and its integrity is protected through the origin AS' signature over the PCB (\cref{sec:pan}).
%
%
Importantly, on-demand RACs execute on-demand algorithms in a sandboxed environment to prevent malicious use; an algorithm's runtime and memory consumption are strictly limited.




\subsection{Egress Gateway}\label{sec:internal_architecture:egress}

\paragraph{PCB Initialization}
When creating new PCBs, the egress gateway of the origin AS extends each PCB with all information that is available and that the AS is willing to share, for example geolocation or latency data.
To benefit from on-demand routing, pull-based routing, or route granularity optimization, the egress gateway needs to add the respective algorithm IDs, a target AS, and interface groups~(\cref{sec:principles}).
The egress gateway then signs the PCB and sends it to the ingress gateway of the corresponding neighboring AS.

\paragraph{PCB Propagation}
When receiving PCBs and egress interface IDs from the RACs, the egress gateway filters them by consulting its \textit{egress database}.
If a received PCB is not yet present in this database, the PCB is inserted together with the egress interface IDs for which it was optimized.
The egress gateway then adds all available path metadata that the AS is willing to share to the PCB.
After signing the PCB, it is propagated to the ingress gateways of the neighboring ASes corresponding to the selected egress IDs.
For pull-based PCBs where the target AS corresponds to the AS of the egress gateway, the PCB is sent back to its origin AS.
If a received PCB is already present in the database, which can happen for PCBs that are optimal according to the computations of multiple different RACs, then the PCB is only propagated on the egress IDs that are newly added to the database.
Similarly to the ingress database, (soon-to-be) outdated PCBs are removed from the egress database on a regular basis.
To reduce the amount of memory needed, the egress database does not store the actual PCBs, but only their hashes.
%



\paragraph{Path Registration}
To make the paths discovered in the PCBs available to end hosts, the egress gateway registers them at the AS' path service.
To improve usability, it tags the PCBs with the set of criteria they were optimized for.
%

\section{Standardization and Deployment}\label{sec:deployment_standardization}
The \system{} architecture, as presented so far, gives individual ASes almost unbounded freedom and flexibility in deploying multiple parallel routing algorithms.
However, routing and path optimization are collaborative efforts, 
and their outcome greatly benefits from having as many ASes as possible participating in the process.
Therefore, the architecture proposed in this paper would not be complete without a \emph{standardization model} to guide ASes towards shared algorithm and implementation choices, such that globally desirable goals are always achieved.

Borrowing from software engineering terminology, we thus propose a tiered standardization model where architectural features are categorized by their criticality to Internet-wide connectivity and their expected stability in time.

\paragraph{Feature Standardization Model}\label{sec:deployment_standardization:criticality}
We identify three standardization tiers for \system{}'s features: 
(i) \emph{stable} features, which are essential to global connectivity and should be changed infrequently, (ii) \emph{beta} features, comprising additional functionality---e.g., optimizing paths on elementary optimality criteria---which is not critical to connectivity, but can greatly benefit from widespread deployment, and finally 
(iii) \emph{nightly} features, which can change extremely frequently---e.g., optimizing paths for arbitrary criteria of specific applications.

We now provide more details on each of these tiers. 

\paragraph{Stable Features}
These features are the foundation of the \system{} architecture, and are ideally standardized once with minimal future updates to maximize their global adoption. The stable features are: 
\begin{itemize}
    \item The basic format of PCBs, compatible with the SCION's legacy PCB format~\cite{Chuat2022SCION}, containing the necessary information to specify paths.
    \item The PCB extensions (\Cref{sec:principles:pcb_extensions}) for  on-demand routing (\Cref{sec:principles:on_demand}), pull-based routing (\Cref{sec:principles:pull_based}), and flexible optimization granularity (\Cref{sec:principles:granularity}).
    \item The interface between RACs (\Cref{sec:internal_architecture:rac}) and algorithms, which needs to be standardized to allow low-effort deployment of new algorithms and to enable on-demand routing. With a standardized interface, algorithms and their implementations can be deployed ubiquitously, favoring adoption and extensibility. 
    \item A single routing algorithm to guarantee global connectivity. While designing an optimal algorithm for fast and reliable global connectivity is beyond the scope of this paper, the default choice may be based on the current SCION routing algorithm.
\end{itemize}

Any AS can participate in \system{} routing as long as it implements these features correctly.

\paragraph{Beta Features}
Elementary optimality criteria---on common path performance metrics such as latency and bandwidth---are likely to be widely adopted and used by many different applications.
Therefore, to promote interoperability and consistent optimization, it is essential to have a standard definition of (i) which metrics are used, (ii) how the metrics are computed, (iii) how each metric is encoded in PCBs, and (iv) which algorithms are globally preferred for each set of elementary criteria. 
Finally, note that extending the list of metrics must be a straightforward task, not to hinder the development of new routing algorithms.
These rapid updates could be supported, for example, by publishing new metrics and algorithms to public append-only lists.
%
ASes can choose whether or not to participate in path optimization on elementary criteria, and if so, with which algorithm. 

\paragraph{Nightly Features}
In this tier, on-demand routing (\Cref{sec:principles:on_demand}) replaces algorithm standardization: This mechanism ensures that all ASes
%
supporting on-demand routing
run the same algorithm, which is specified by the origin AS within PCBs.
Routing algorithms disseminated in this way are not standardized, allowing origin ASes to freely choose any algorithm they want to deploy. 
These algorithms use the publicly-available \emph{beta} metrics.

\section{Implementation and Evaluation}\label{sec:implementation}
In this section, we demonstrate \system{}'s practical viability.
Parties interested in our research can get access to our implementation, add new RACs, and run experiments using their own algorithms with minimal effort.\footnote{Source code available at \href{https://anonymous.4open.science/r/scion-irec/}{https://anonymous.4open.science/r/scion-irec/}}

\subsection{Implementation}
We implement \system{} in Go based on the open-source SCION codebase~\cite{scionproto}, where we substitute the legacy SCION control service with \system{}'s intra-AS components (\Cref{sec:internal_architecture}).
The ingress gateway, the egress gateway, and the RACs are implemented as separate processes; if required, all these components plus the ingress and egress databases can be run on separate machines.
Still, to facilitate deployment we further implement an alternative version that aggregates the functionality of both gateways and databases into a single Go process.
Communication between the processes is implemented using gRPC remote procedure calls~\cite{gRPC}, and we use SQLite~\cite{SQLite} for the ingress- and egress databases.
To ensure interoperability with neighboring ASes not supporting \system{}, the ingress gateway listens on the same service address as the globally standardized service address of the legacy control service~\cite{scion-control-service}.

\paragraph{RACs} To reduce maintenance effort and cost, we develop a unified implementation for static and on-demand RACs, where the type is specified in a RAC's configuration file upon initialization.
Both static and on-demand routing algorithms are compiled as WebAssembly\cite{WebAssembly} modules.
WebAssembly code is created by compiling a higher-level programming language into WebAssembly bytecode, allowing algorithm developers to write code in familiar languages. 
To sandbox algorithms, RACs use Wasmtime~\cite{Wasmtime}, a standalone runtime for running WebAssembly code.
Parameter values passed by a RACs to algorithms are (i) PCBs marshalled using Protobuf~\cite{Google:2023:Protobuf}, (ii) a function to request information about the AS topology, and (iii) functions to return the per-egress-interface optimal sets of PCBs to the RAC.
Flexible path optimization (\Cref{sec:principles:granularity}) and pull-based routing (\Cref{sec:principles:pull_based}) are optional features that can be turned on or off for each RAC independently.

\subsection{Evaluation}
In two separate experiments, we evaluate the PCB processing latency and throughput of (i) our RAC implementation, configured as an on-demand RAC (i.e., the one with higher overhead), and the latency of (ii) the legacy SCION control service implementation. 
We use the legacy SCION routing algorithm, which selects the \num{20} shortest paths from each origin AS, and benchmark the implementations with a set~$\Phi$ of candidate PCBs of various sizes.
Once the algorithms has computed the set of optimal PCBs from $\Phi$, the RAC immediately fetches $\Phi$ and runs the algorithm again.
We run our experiments on a machine with two \num{18}-core Intel Xeon E5-2695~\cite{Intel:2023:Xeon} processors and \SI{128}{\giga\byte} of memory.



\paragraph{Latency}
\Cref{fig:impl:latency} shows the processing latency of \system{} compared to the legacy control service, for different sizes of~$\Phi$.
The total latency for \system{} comprises the setup of the Wasmtime environment, the gRPC calls, and executing the algorithm in form of a WebAssembly module.
For $|\Phi|=64$, \system{}'s latency is \textasciitilde\num{426} times higher than the one of the legacy control service.
If the candidate PCB set is smaller than the desired number of optimal PCBs, which is \num{20} in our experiment, the algorithm can immediately return.
This characteristic makes the legacy control service exceptionally efficient for candidate sets containing fewer than \num{20} beacons.
Still, for both versions, i.e., \system{} and legacy SCION, the latency is negligible compared to the PCB propagation interval.



\begin{figure}
    \centering
    \includegraphics[width=\linewidth]{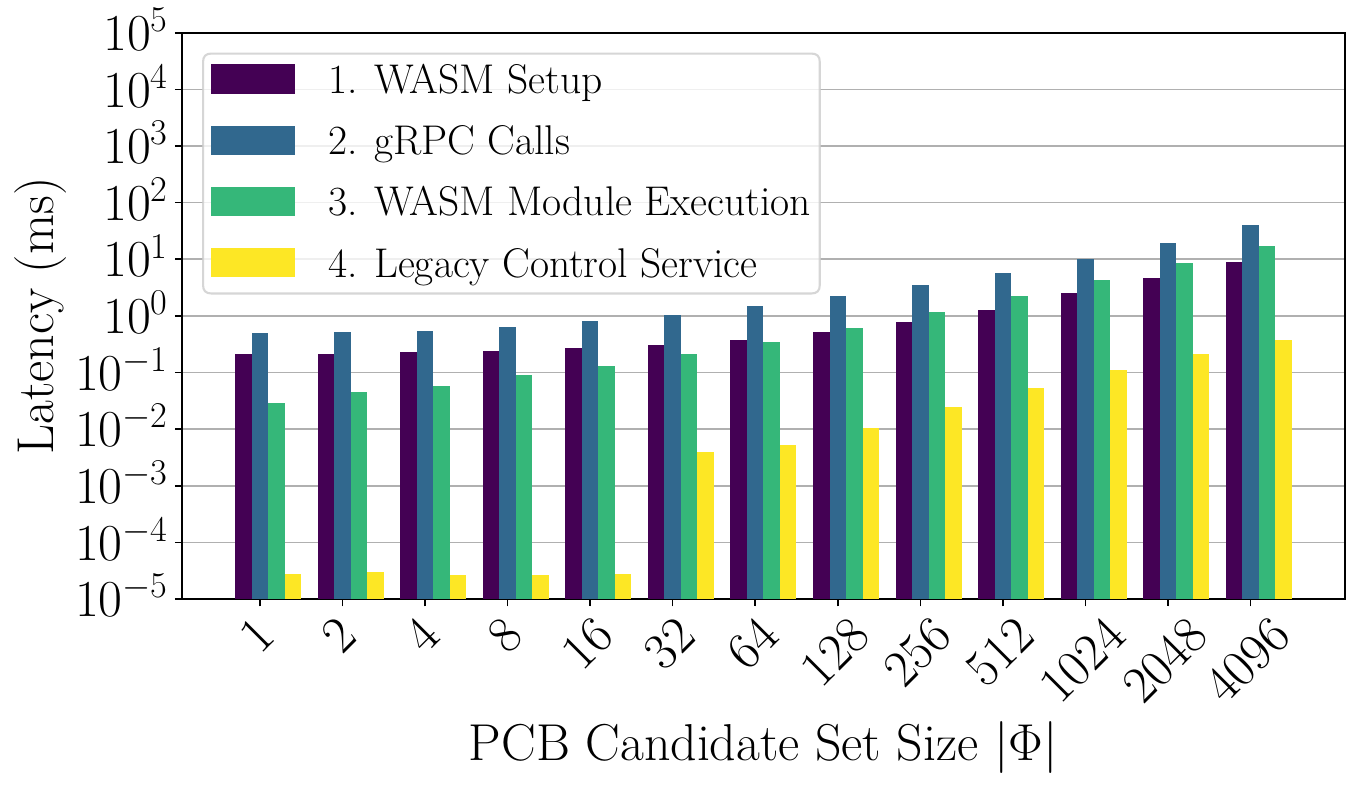}
    \caption{Processing latency for different \system{} sub-tasks (1-3) compared to the legacy SCION control service (4), for varying sizes of the PCB candidate set~$\Phi$.}
    \label{fig:impl:latency}
\end{figure}

\paragraph{Throughput}
\Cref{fig:impl:throughput} shows the PCB processing throughput of \system{} for different numbers of RACs. 
We observe that the throughput increases linearly with the number of RACs, indicating a close-to-optimal scalability; it increases sub-linearly with the size of~$\Phi$.
The execution time of the WASM algorithm approximately doubles when the size of $\Phi$ doubles, while the WASM setup and gRPC do not exhibit the same doubling. As a result, for larger $\Phi$ the per-beacon overhead decreases proportionally, leading to a higher throughput.
The throughput can be further improved by running more RACs on additional machines.

\paragraph{Backward Compatibility}
In addition to the latency and the throughput measurements, we evaluate our implementation by creating and attaching our own SCION AS to the global SCIONLab research network~\cite{Kwon:2020:SCIONLab}.
We observed that \system{} is fully backward compatible with SCIONLab's legacy ASes, with no interruptions in connectivity.
\system{} can thus be deployed incrementally at the level of individual ASes, without requiring global coordination.


\begin{figure}
    \centering
    \includegraphics[width=\linewidth]{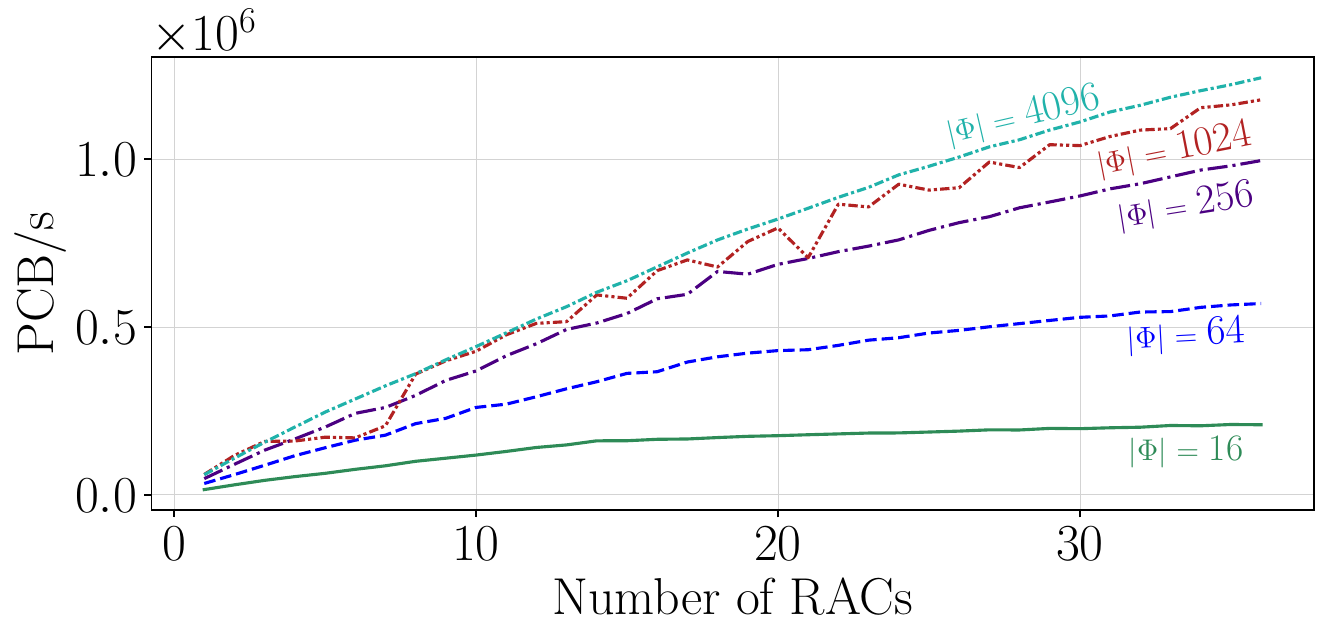}
    \caption{
    PCB processing throughput for different number of \system{} RACs, for varying sizes of the PCB candidate set~$\Phi$. 
    }
    \label{fig:impl:throughput}
\end{figure}
\section{Large-Scale Simulations}\label{sec:simulations}

To fully demonstrate \system{}'s path-optimization potentials, we need to evaluate it on a realistic Internet-scale topology with massive path diversity. To achieve this in the SCION production network or in SCIONLab~\cite{Kwon:2020:SCIONLab}, we would need to convince hundreds of AS operators, which is infeasible for research purposes. Therefore, we rely on simulations.\footnote{Source code at \href{https://anonymous.4open.science/r/scion-irec-simulations-3F82/}{https://anonymous.4open.science/r/scion-irec-simulations-3F82/}}

\subsection{Simulation Setup and Topology}\label{sec:simulations:setup}

We implement \system{} in the ns-3-based~\cite{nsnam_ns-3_nodate} SCION simulator~\cite{scion-sim} running on a machine with two \num{64}-core AMD EPYC~\cite{AMD:2023:EPYC} processors and \SI{512}{\giga\byte} of memory.
To create the simulation topology, we use a subset of the CAIDA geo-rel dataset~\cite{CAIDA-Data-Geo}, which includes relationships between ASes and the locations of the inter-AS links.
The location of inter-domain links allows estimating the propagation delay between border routers of each AS using the great circle distance.
We iteratively prune the lowest-degree ASes from the dataset to arrive at a topology consisting of the \num{500} highest-degree ASes and more than \num{100000} inter-domain links.

\subsection{Implemented Mechanisms and Algorithms}
We implement \system{} and simulate it with four static RACs and one on-demand RAC in each AS (\Cref{sec:internal_architecture:rac}). All RACs optimize and propagate PCBs periodically every ten simulated minutes. 
We limit the number of paths that can be registered at the path service per RAC, origin AS, and interface group to \num{20}. We base our evaluation on the registered paths only, i.e., the ones available to endpoints.

\paragraph{Static RACs} In an AS, each of the following four algorithms runs on a separate static RAC and optimizes on a specific criterion that takes into account performance metrics available in our topology: (i) \emph{shortest path (1SP)}, optimizing on the AS-level path length by propagating the shortest path for each origin AS on all egress interfaces, (ii) \emph{five shortest paths (5SP)}, optimizing path length by propagating the five-shortest paths on all egress interfaces, (iii) \emph{heuristic disjointness (HD)}, proposed by Krähenbühl et al.~\cite{conext2021deployment}, heuristically optimizing inter-domain link disjointness, and (iv) \emph{delay optimization (DO)}, optimizing the propagation delay of paths calculated by accumulating the estimated great-circle delays of all on-path AS hops. We consider different versions of this algorithm, as follows.

We jointly evaluate flexible optimization granularity (\Cref{sec:principles:granularity}) and optimization on extended paths (\Cref{sec:principles:per_egress}) as both mechanisms tackle the impact of ASes' internal network on optimality of inter-domain paths. As the internal network of an AS cannot affect AS-level path length or disjointness of inter-domain links, we evaluate the effect of these mechanisms on the performance of the delay optimization algorithm. We design two experiments: (i) \emph{DON}, i.e., DO with \textit{none} of the mechanisms applied, and (ii) \emph{DOB}, i.e., DO with \textit{both} mechanisms applied. We define interface groups based on the routers' geographic locations, suitable for delay optimization, and investigate the effect of different interface group granularities, i.e., we evaluate two configurations of interface groups with distance between any two interfaces no more than \SI{300}{\kilo\meter} (\emph{DOB300}) and \SI{2000}{\kilo\meter} (\emph{DOB2000}), respectively.

\paragraph{On-Demand RACs} To show the benefits of on-demand (\Cref{sec:principles:on_demand}) and pull-based routing (\Cref{sec:principles:pull_based}), we design the \emph{pull-based disjointness (PD)} algorithm that uses those mechanisms to achieve path disjointness unattainable by algorithms that cannot make use of those mechanisms.
The algorithm allows an AS to iteratively construct a set of link-disjoint paths to any target AS by starting from a non-empty set of paths to the target AS, already discovered by other algorithms; we use HD in our setup. In each iteration, the AS originates on-demand pull-based PCBs, specifying the target AS and a new algorithm that avoids PCB propagation on links in the set of paths to the target AS.
When some of these PCBs ultimately arrive at the target AS, it returns them to the origin AS, which only adds the first-received PCB of the iteration to its set and starts the next iteration. 
The AS repeats this process until it discovers desired number of disjoint paths, i.e., in \num{20} our setup.

\subsection{Results}\label{sec:simulations:results}
\begin{figure*}[h!]
    \begin{subfigure}[t]{0.32\linewidth}
        \centering
        \includegraphics[width=\linewidth]{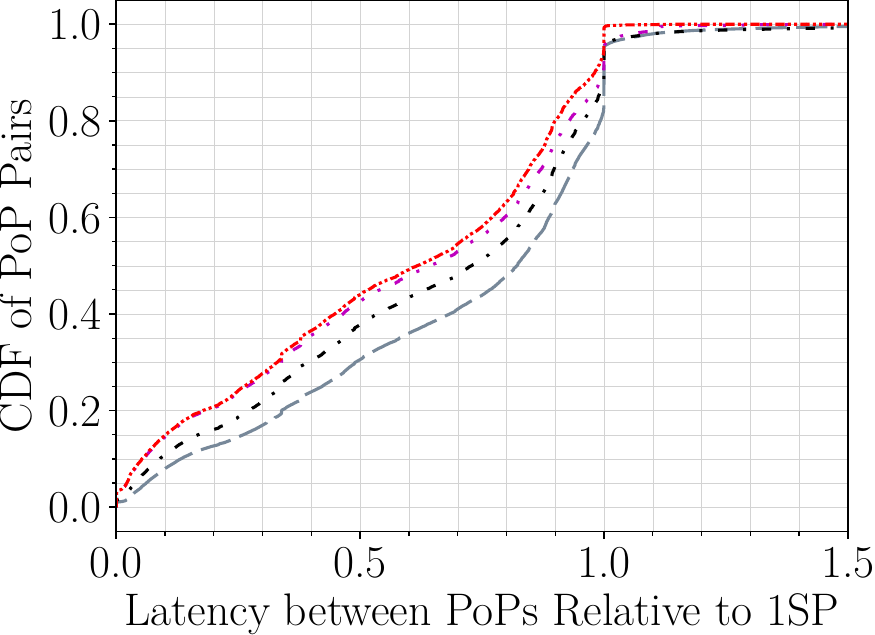}
        \caption{}
        \label{fig:simulations:latency}
    \end{subfigure}\hfill
    \begin{subfigure}[t]{0.32\linewidth}
        \centering
        \includegraphics[width=\linewidth]{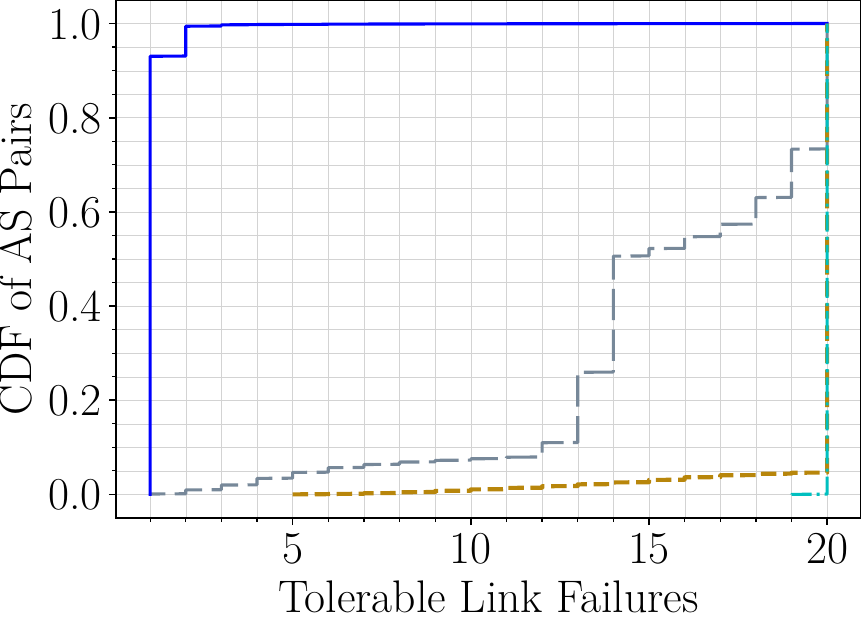}
        \caption{}
        \label{fig:simulations:tolerable}
    \end{subfigure}\hfill
    \begin{subfigure}[t]{0.32\linewidth}
        \centering
        \includegraphics[width=\linewidth]{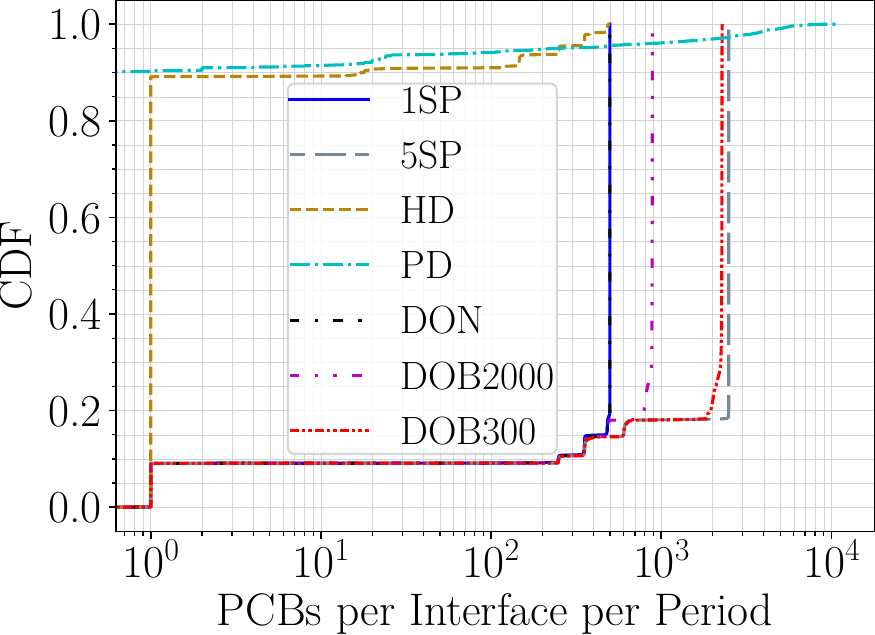}
        \caption{}
        \label{fig:simulations:pcb_counts}
    \end{subfigure}
    \caption{\system{}'s simulation results expressed as cumulative distribution functions (CDFs) for a topology of \num{500} ASes.}
    \label{fig:simulations:simulation_results}
\end{figure*}

\paragraph{Propagation Delay} We define a point of presence (PoP) of an AS as a geolocation where it has at least one inter-domain link, and evaluate the minimum propagation delay each algorithm can provide between any pair of PoPs in two \emph{different} ASes. If an algorithm cannot find a direct inter-domain path between a specific pair of PoPs in two ASes, we take into account available inter-domain paths between other PoPs of the same AS pairs and add the \emph{intra-domain} great-circle delay between end PoPs of paths and the desired PoPs.

The distribution of minimum delays for 5SP, DON, DOB300, and DOB2000 relative to 1SP are illustrated in~\Cref{fig:simulations:latency}. We observe that these algorithms significantly reduce the delay for most of the PoP pairs. Among them, DO variants and specifically DOB variants show the lowest delays. In particular, DOB300 provides the lowest delay,  \SI{20}{\%} and \SI{10}{\%} lower delay than 5SP, and DON, respectively, for at least half of PoP pairs.
The greater-than-one tails in the figure correspond to PoP pairs for which 1SP finds an inter-domain path, while other algorithms do not.
DOB300 shows significantly fewer such PoP pairs because of more fine-grained interface groups, increasing the number of PoP pairs with inter-domain paths connecting them.

\paragraph{Disjointness} As a metric for disjointness, we define \emph{tolerable link failures (TLF)} between a pair of ASes as the minimum number of links on discovered paths that can be removed until all those paths are disconnected. 
\Cref{fig:simulations:tolerable} depicts the TLF distribution for 1SP, 5SP, HD, and PD.
1SP and 5SP achieves the maximum TLF only for \SI{1}{\%}, and \SI{15}{\%} of AS pairs, respectively. In contrast, HD achieves this goal for more than \SI{95}{\%} of AS pairs, and the PD achieves it for almost all AS pairs. Note that since each algorithm has a limit of \num{20} paths to register at the path server, the theoretical maximum TLF is \num{20}, when all paths are completely disjoint.

\paragraph{Propagated PCBs}
\Cref{fig:simulations:pcb_counts} shows the distributions of the number of PCB sent by different algorithms over all interfaces and periods. The number of propagated PCBs provides a metric for message complexity and communication overhead. We observe that 1SP, 5SP, DON, DOB2000, and DOB300 show similar patterns, because all of them uniformly propagate PCBs on all egress interfaces. More specifically, their overhead does not vary for \SI{80}{\%} of interfaces, and among these algorithms, 5SP has the highest and 1SP has the lowest overhead, respectively. Furthermore, both DOB-variants' overhead grows with the number of interface groups in the whole network. In contrast, PD and HD show a different pattern with low overhead during most periods. This is because interfaces on which PCBs have been propagated before are avoided in subsequent periods.
However, PD shows a high overhead during few periods, which is because it discovers paths per pair of ASes instead of per origin AS.

\section{Discussion and Future Work}\label{sec:discussion}

%

\paragraph{End-to-End Performance}
Between a server and different border routers of its AS, a metric's performance might vary significantly.
Even if an end host selects the best inter-domain path towards this AS, the end-to-end path to the server might therefore not be optimal.
In \system{}, PCBs do not carry the necessary information for every server in the AS due to the possibly large communication overhead.
Instead, for measurable metrics like latency or bandwidth, the client can discover the optimal end-to-end path simply by connecting to the server over paths traversing interfaces belonging to different interface groups.
In the future, DNS information could be extended with metric information regarding its recommended interface group.

\paragraph{Bootstrapping Connectivity}
Disconnected ASes, as well as newly joining ASes, want to quickly (re-)establish connectivity.
This includes (i) finding paths to other ASes and (ii) propagating self-originated PCBs throughout the Internet.
\system{} can be extended to achieve this on the order of a \textit{single round trip}.
First, for paths to other ASes, the ingress gateway of the disconnected AS can request paths from the egress gateways of its neighboring ASes.
In case the neighboring ASes also do not have any paths available, the process continues recursively.
Second, to achieve rapid PCB propagation, an ingress gateway notifies a dedicated RAC immediately when new PCBs arrive, where this RAC forwards the PCBs directly to the egress gateway.
Therefore, already the first PCB sent by a (re-)connecting AS is propagated instantaneously through the whole Internet.
To ensure scalability of this approach, the RAC only guarantees each origin AS to forward a single (potentially sub-optimal) PCB per certain time interval.

\section{Related Work}\label{sec:related}

\paragraph{Extensible Inter-Domain Routing}
XIA~\cite{Naylor:2014:XIA}, Trotsky~\cite{McCauley2019Trotsky}, and xBGP~\cite{Wirtgen:2023:xbgp} take significant steps towards an extensible Internet. XIA proposes an expressive Internet architecture with native support for multiple principals and the ability to evolve to accommodate new ones. Trotsky enables the introduction of new designs and Internet architectures through its backward-compatible framework, and xBGP~\cite{Wirtgen:2023:xbgp} enables BGP to extend by introducing user-defined and dynamically addable eBPF bytecodes.
Thus, network operators can provide and deploy their own extensions without waiting for standardization or vendors. However, they do not consider multi-criteria path optimization as a design objective and do not provide end domains with means of expressing their diverse criteria to the control plane. In particular, in xBGP each AS decides its routing policy and optimization criteria independently, not providing means for global optimality guarantees for different criteria. Furthermore, each AS can optimize paths on a single set of criteria at any given time, and is oblivious to the criteria of endpoints outside of its domain.

\paragraph{Multi-Criteria Path Optimization}
Multi-objective path problems consider path performance metrics as Cartesian product of elementary metrics, extending either by addition (e.g., latency), or by min/max (e.g., capacity), and order paths partially using the product order of their term-wise total orders~\cite{Brumbaugh1989empirical, Hansen1980Bicriterion, MARTINS1984236}. These problems, however, consider only a limited category of performance metrics.
Sobrinho et al.~\cite{Sobrinho:2020:multiple_optimality} guarantee optimality on multiple criteria by defining a partial order on the intersection of all criteria and selecting the set of dominant paths, i.e., the ones to which no other path is preferable and which are incomparable among each other.
However, the number of incomparable paths grows significantly by increasing the number of criteria in the intersection, resulting in significant communication cost.
Furthermore, their method neglects extensibility as it complicates the optimization logic, requires it to be deployed by all ASes, and requires its interruption and modification to add any new criterion.

\paragraph{On-Demand Routing} Yampolskiy et al.~\cite{Yampolskiy2010multidomain} propose a resource reservation algorithm to establish connections with multiple QoS constraints per endpoints' requests in inter-domain networks. Route Bazaar~\cite{route_bazaar_2015} introduces a system enabling customers and ASes to agree on routes according to QoS criteria. These methods have two fundamental differences with \system{}: (i) their path discovery is accompanied with a QoS agreement between ASes, while \system{} is a general purpose routing architecture, and (ii) they route on routing constraints, i.e., to satisfy thresholds for performance metrics, while \system{} optimizes paths, i.e., discovers the best possible path according to criteria, and it can also satisfy constraints through on-demand and pull-based routing.


%

\section{Conclusion}\label{sec:conclusion}

Despite tremendous advancements witnessed in the realm of information technology over the past decades, along with a
continuous increase in dependence on communication networks, inter-domain
routing has endured a conspicuous lack of transformation over the course of  the past \num{25} years.
In particular, despite the ever-increasing diversity of applications' communication requirements, BGP has largely remained static over the past 25 years.

\system{} overcomes those limitations with its extensible control-plane architecture that enables routing on many criteria.
By optimizing paths on independent sets of criteria using parallel routing algorithms, \system{} achieves extensibility and many criteria path optimization.
Based on this idea, we also introduce other mechanisms that allow real-time extensibility, guarantee optimality, and enable both end ASes of traffic to express their path optimality criteria to the control plane.

Our implementation in SCION and large-scale simulations based on ns-3 are open source and allow researchers to run and evaluate their own algorithms in \system{}.
In this paper, considering propagation delay and link disjointness as optimization criteria, we reduce propagation delay by at least \SI{40}{\milli\s} for \num{25}\% of PoP pairs and achieve significant path disjointness with overheads comparable to conventional algorithms.

\system{} opens up new opportunities for exciting research in inter-domain routing, ultimately leading to enhanced communication quality for endpoints and applications---all in real-time, with negligible effort, and no interruption in communication.


\clearpage

\bibliographystyle{ieeetr}
\bibliography{rfc}


\end{document}